\documentclass[ prl, twocolumn, superscriptaddress, amsfonts, amsmath,floatfix]{revtex4-1}
\pdfoutput=1
\usepackage{graphicx}
\usepackage{amssymb,amsmath}
\usepackage{xspace}
\usepackage[usenames, dvipsnames]{color}
\usepackage{natbib}

\newcommand{\be}{\begin{equation}}
\newcommand{\ee}{\end{equation}}

\newcommand{\rg}{{\bf r}}

\newcommand{\Eg}{{\bf E}}
\newcommand{\Bg}{{\bf B}}
\newcommand{\Dg}{{\bf D}}
\newcommand{\Mg}{{\bf M}}

\begin{document}

\title{Universal statistics of waves in a random time-varying medium}

\author{R. Carminati} 
\altaffiliation{remi.carminati@espci.psl.eu}
\affiliation{Institut Langevin, ESPCI Paris, PSL University, CNRS, 1 rue Jussieu, F-75005 Paris, France}
\author{H. Chen}
\affiliation{Institut Langevin, ESPCI Paris, PSL University, CNRS, 1 rue Jussieu, F-75005 Paris, France}
\author{R. Pierrat}
\affiliation{Institut Langevin, ESPCI Paris, PSL University, CNRS, 1 rue Jussieu, F-75005 Paris, France}
\author{B. Shapiro}
\affiliation{Department of Physics, Technion-Israel Institute of Technology, Haifa IL-32000, Israel}

\begin{abstract}
We study the propagation of waves in a medium in which the wave velocity fluctuates randomly in time. We prove that at long times, the statistical distribution of the wave energy is log-normal, with the average energy growing exponentially. For weak disorder, another regime preexists at shorter times, in which the energy follows a negative exponential distribution, with an average value growing linearly with time. The theory is in perfect agreement with numerical simulations, and applies to different kinds of waves. The existence of such universal statistics bridges the fields of wave propagation in time-disordered and space-disordered media.
\end{abstract}

\maketitle

{\it Introduction.}
In recent years there has been a growing interest in space-time metamaterials for electromagnetic~\cite{Caloz2020} or acoustic waves~\cite{Fleury2019}. These are materials whose properties are modulated in space and time. Homogeneous materials modulated only in time, referred to as ``temporal" or ``pure-time", offer new approaches for the control of waves, {\it e.g.} through the design of active metasurfaces. They also stimulate basic studies in wave physics. For example, it is known for electromagnetic waves that when the dielectric function is suddenly changed from one value to another, a backward propagating (time-reversed) wave appears~\cite{Morgenthaler1958}. This phenomenon has been recently put into a general framework, and demonstrated experimentally with water waves~\cite{Bacot2016,Bacot2019}. A periodic modulation of the dielectric function has also been investigated~\cite{Holberg1966,Zurita2009}, leading to the appearance of bands and gaps in the wave propagation constant $k$, as well as topological phases~\cite{Lustig2018}. Building on the analogy between space and time, new approaches have been proposed for the control of waves~\cite{Caloz2018,Pacheco2020,Pacheco2020b}. There is a lot to expect in the interaction between waves and more complex temporal materials, including random time-varying media, a domain that has remained unexplored to a large extent.

The purpose of this Letter is the study of wave propagation in a medium with a dielectric function $\varepsilon(t)$ fluctuating randomly in time. Our focus is on electromagnetic waves, but the developed theory and the results encompass other kind of waves, such as acoustic or water waves. The question of time evolution of a pulse subjected to random``kicks'', due to sudden changes in $\varepsilon(t)$, has been posed and studied in Ref.~\cite{Sharabi2021}. It has been shown that, after a sufficiently long time, the energy $U(t)$ of the pulse increases exponentially. A similar regime has been found in a recent study, in which water waves propagate in a disordered time-periodic lattice~\cite{Apffel2021}. This behavior suggests a connection with Anderson localization of waves in a spatially disordered medium~\cite{Anderson}. Here we consider a general model of a disordered time-varying medium, with an emphasis on weak disorder (the criterion for weak disorder will be stated later). In this case, one can develop a detailed analytical theory. The theory shows that at times larger than a crossover time $\tau_c$, $\langle \ln U(t) \rangle$ becomes proportional to $t$, with the brackets denoting the average value, in agreement with experimental and numerical observations~\cite{Sharabi2021,Apffel2021}. Interestingly, there is an intermediate regime $\tau_m \ll t \ll \tau_c$, with $\tau_m$ the microscopic time (defined as the typical time of the modulation of $\varepsilon(t)$), in which the average energy $\langle U(t) \rangle$ grows linearly with $t$. The full statistical distribution of the energy $U$ can be determined in both regimes. In the intermediate regime, the energy follows a negative exponential distribution. For long times $t \gg \tau_c$, the statistics becomes log-normal, in agreement with known results in one-dimensional wave transport.

{\it General framework.}
We consider the propagation of electromagnetic waves in a homogeneous, isotropic and nonmagnetic medium, described by its time-dependent dielectric function $\varepsilon(t)$ such that the displacement and electric fields are related by $\Dg(\rg,t) = \varepsilon_0 \varepsilon(t) \Eg(\rg,t)$. When the displacement field has a single component $D$, and depends only on one space coordinate $x$, it satisfies
\begin{equation}
\frac{\partial^2 D}{\partial x^2}(x,t) - \frac{\varepsilon(t)}{c^2} \frac{\partial^2 D}{\partial t^2}(x,t) = 0 \, ,
\label{eq:wave_1D}
\end{equation}
together with appropriate boundary conditions. It will prove useful to perform the analysis in $k$ space. The space Fourier transform, defined as $D(k,t) = \int_{-\infty}^{+\infty} D(x,t)  \exp(-ikx) dx$, satisfies
\begin{equation}
\frac{\partial^2 D}{\partial t^2}(k,t) + \Omega^2(t) D(k,t) = 0 \, ,
\label{eq:Helmholtz_k_space}
\end{equation}
where $\Omega^2(t) = c^2 k^2/\varepsilon(t)$. Equation~(\ref{eq:Helmholtz_k_space}), supplemented with two initial conditions for $D(k,t)$ and its time derivative, constitutes a Cauchy problem. We note that since $D(x,t)$ is real, $D(-k,t)=D^*(k,t)$, where the superscript $*$ stands for complex conjugate. Therefore the analysis can be limited to $k \geq 0$. {We also point out that this description is not limited to a fully homogeneous space. Indeed, the only requirement is homogeneity along the propagation direction $x$. For example, the analysis could apply to a waveguide filled with a homogeneous medium having $\varepsilon(t)$ depending on time, with the plane wave replaced by a guided wave with a given transverse profile.}

Over a time interval in which $\varepsilon$ is a constant, the general solution to Eq.~(\ref{eq:Helmholtz_k_space}) is of the form
\begin{equation}
D(k,t) = D^+(k,t) + D^-(k,t) \, ,
\label{eq:solution_Helm_Fourier}
\end{equation}
with $D^+(k,t) \sim \exp(-i\Omega t)$ and $D^-(k,t) \sim\exp(i\Omega t)$, corresponding to plane waves propagating in the positive and negative $x$ direction, respectively. In this study, the observable of interest is the electromagnetic energy 
$U(t) =[2 \varepsilon_0 \varepsilon(t)]^{-1} \int \Dg^2(\rg,t) d^3 r +  (2 \mu_0)^{-1} \int \Bg^2(\rg,t) d^3 r$~\cite{Landau_Book}.
For a one-dimensional and linearly polarized field, the energy can be rewritten as
\begin{equation}
U(t) = \frac{1}{2\pi\varepsilon_0 \varepsilon} \int_{-\infty}^{+\infty} \left [ |D^+(k,t)|^2 +  |D^-(k,t)|^2 \right ] dk  \, .
\label{eq:energy_def}
\end{equation}
Note that the electric and magnetic contributions to the energy contain interference terms that exactly compensate, resulting in the simple expression above.

\begin{figure}[h]
     \begin{center}
     \includegraphics[width=8cm]{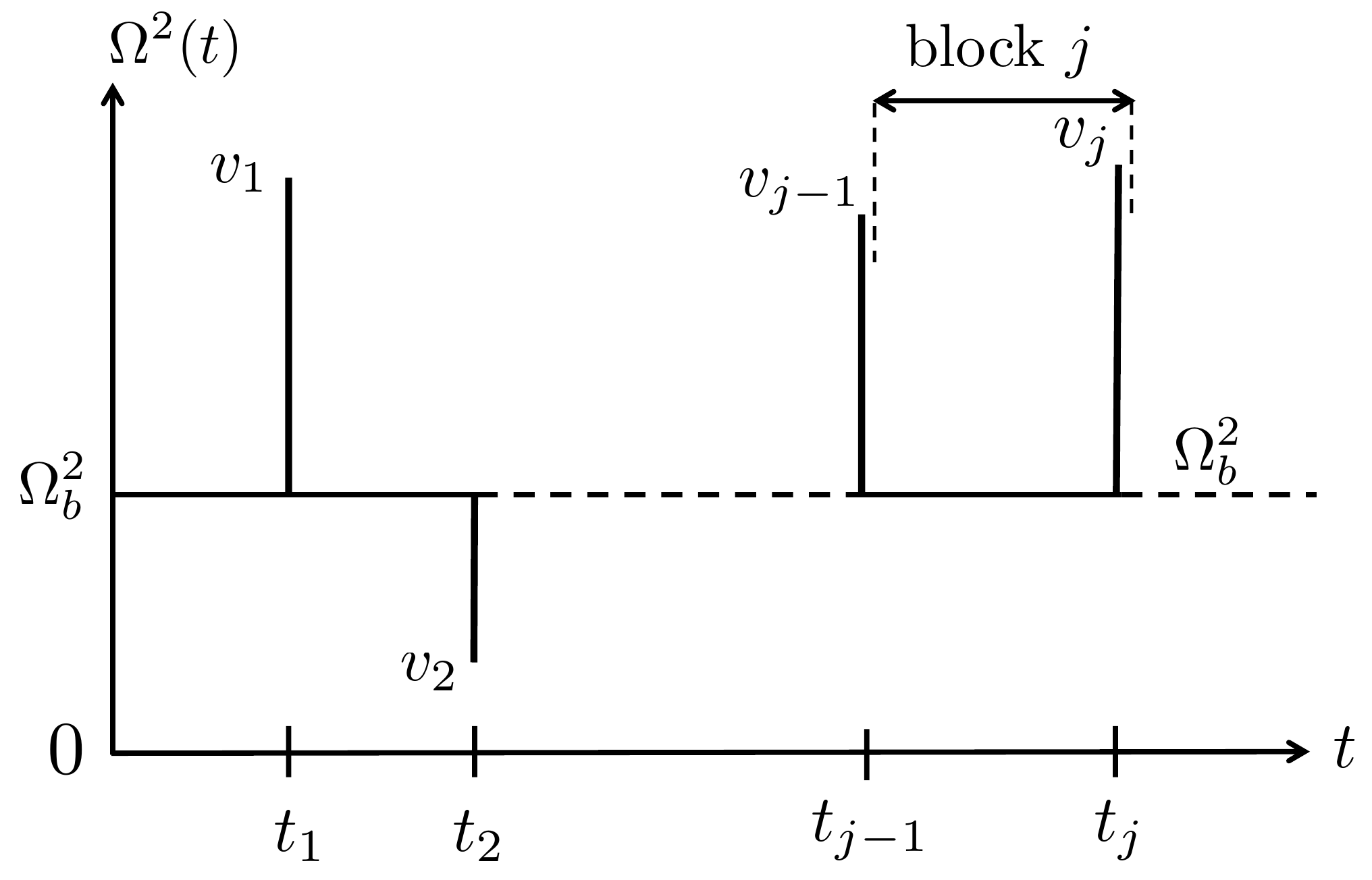}
     \end{center}    
     \caption{\label{fig:delta_kicks} Random chain of $\delta$ kicks, with $\Omega^2(t) = \Omega_b^2 + \sum_j v_j \, \delta(t-t_j)$. The kick strengths $v_j$ and the times $t_j$ are random variables. After each kick the medium recovers the background value $\Omega_b$. The statistical distribution of $v_j$ is independent of $j$.
       }
\end{figure}

{\it Transfer matrix.}
In $k$ space, the time evolution of the field can be described using transfer matrices. 
To model a medium with a dielectric function $\varepsilon(t)$ changing randomly in time, we can take {$\Omega^2(t)$ to be a series of instantaneous kicks ($\delta$-kicks) on top of a background value $\Omega^2_b$,} as represented in Fig.~\ref{fig:delta_kicks}. In this model the kick strength $v_j$ and times $t_j$ are independent random variables.

The transfer matrix $\Mg_j$ of a single block $j$ (see Fig.~\ref{fig:delta_kicks}) connects the field $D_j$ at time $t_j + 0$ to the field $D_{j-1}$ at time $t_{j-1}+0$, in such a way that
\begin{equation}
\begin{bmatrix}
D^+_j \\
D^-_j
\end{bmatrix}
=
\Mg_j
\begin{bmatrix}
D^+_{j-1} \\
D^-_{j-1}
\end{bmatrix} \, .
\label{eq:def_Mj}
\end{equation}
{The analysis is performed at a fixed $k$, and the matrix elements depend on $k$, but we drop the argument $k$ in the notations for simplicity.}
An explicit calculation of the transfer matrix, {detailed in the Supplemental Material~\cite{SM}}, shows that
\begin{equation}
\Mg_j
=
\begin{bmatrix}
a_j & b_j \\
b_j^* & a_j^*
\end{bmatrix}
\label{eq:Mj_general}
\end{equation}
with
\begin{eqnarray}
a_j &=&  (1-iu_j) \exp(-i\theta_j) \nonumber \\
b_j &=& -iu_j \exp(i\theta_j) \, ,  \label{eq:ajbj_kicks}
\end{eqnarray}
and $u_j=v_j/(2 \Omega_b)$. The matrix $\Mg_j$ has the following properties:
\begin{align}
&|a_j|^2 \geq 1
\label{eq:property1} \\
&\mathrm{det} \Mg_j = |a_j|^2 - |b_j|^2 = 1 \, .
\label{eq:M_unimodular}
\end{align}
Moreover, due to property (\ref{eq:M_unimodular}), it is easy to see that $|D_j^+|^2 - |D_j^-|^2 = |D_{j-1}^+|^2 - |D_{j-1}^-|^2$, showing that the quantity $|D_j^+|^2 - |D_j^-|^2$ is conserved. At time $t<t_1$ (before the first kick), we can assume that the incident wave propagates in the positive $x$ direction, with $D_0^+$ normalized such that $|D_0^+|^2=1$. As a result, the following relation is satisfied for all $j$:
\begin{equation}
|D_j^+|^2 - |D_j^-|^2 = 1 \, .
\label{eq:flux_conservation}
\end{equation}
It is important to note that the subsequent analytical treatment is not limited to the specific shape of modulation, like the $\delta$ kicks in Figs.~\ref{fig:delta_kicks}, but is applicable to any modulation provided that $\Omega(t)$ recovers the same background value $\Omega_b$ between the random kicks. {For example we show in the Supplemental Material~\cite{SM} that a model based on rectangular time barriers leads to a transfer matrix satisfying the same properties.} In fact the transfer matrix (\ref{eq:Mj_general}), with the properties (\ref{eq:property1}-\ref{eq:M_unimodular}) for its matrix elements, is the most general $2\times2$ transfer matrix~{\cite{MelloBook}}. 

{\it Statistical theory.}
We now develop a theoretical analysis of the statistical properties of the quantity
\begin{equation}
U_j =|D_j^+|^2 + |D_j^-|^2 \, ,
\label{eq:energy_Uj}
\end{equation}
which, up to a prefactor that we omit, corresponds to the energy in the field after $j$ kicks. It will prove useful to introduce new variables $z_j = |D_j^-|^2$ and $\beta_j = |b_j|^2$. We note that due to relations (\ref{eq:M_unimodular}) and (\ref{eq:flux_conservation}), we have $U_j = 1+2z_j$ and $|a_j|^2 + |b_j|^2=1+2\beta_j$. From Eqs.~(\ref{eq:def_Mj}) and (\ref{eq:Mj_general}), we find that
\begin{align}
\ln(1+2z_j) = \ln(1+2z_{j-1}) + \ln(1+2\beta_j) \nonumber \\
+ \ln \left (1 + 2 \gamma_j \, \frac{\sqrt{z_{j-1}(1+z_{j-1})}}{1+2z_{j-1}} \, \cos \Theta_j \right ) \, ,
\label{eq:recurrence_general}
\end{align}
where $\Theta_j$ is the cumulative phase such that $a_jb_j^* D_{j-1}^+ D_{j-1}^{-*} = |a_j| |b_j| |D_{j-1}^+| |D_{j-1}^-| \exp(i\Theta_j)$ and $\gamma_j=2\sqrt{\beta_j(1+\beta_j)}/(1+2\beta_j)$. The recursion relation (\ref{eq:recurrence_general}) is our starting point in the analysis of the statistical properties of the field energy. 

We first analyze the behavior of the energy $U_N$ after $N$ kicks, in the limit $N \to \infty$. {In this limit, which is the same as $t\to\infty$, the field amplitude $D(k,t)$ is expected to increase exponentially with time. A more rigorous statement is that the Lyapunov exponent
\begin{equation}
\lambda(k)=\lim_{t\to\infty} \frac{\ln |D(k,t)|}{t}
\label{eq:Lyapunov}
\end{equation}
takes a finite positive value. The Lyapunov exponent is a self-averaged quantity, independent of the particular realization of disorder.}
It is also related to the larger eigenvalue $\nu(k,N)$ of the transfer matrix $M(N)=M_N*M_{N-1}...M_1$ corresponding to a chain of $N$ random kicks. Actually $\lambda(k) = \lim_{N\to\infty}\ln\nu(k,N)/N$~\cite{Comtet2013}, which in terms of the energy $U_N$ reads as $\lambda(k) = \lim_{N\to\infty}\ln U_N/2N$. Thus, for large $j$, the variable $z_j$ is exponentially large and we can assume
\begin{equation}
2\sqrt{z_{j-1}(1+z_{j-1})}/(1+2z_{j-1}) \simeq 1 \, ,
\label{eq:condition2}
\end{equation}
which allows us to simplify Eq.~(\ref{eq:recurrence_general}) into
\begin{equation}
\ln U_j = \ln U_{j-1} + \ln(1+2\beta_j) + \ln \left (1 + \gamma_j \,  \cos \Theta_j \right ) \, .
\label{eq:recurrence1}
\end{equation}
We now introduce the important assumption that the cumulative phase $\Theta_j$ is completely random, with a uniform distribution over $[0,2\pi]$. This assumption defines the so-called ``random phase'' model. It cannot be generally valid for modulations like those in Fig.~\ref{fig:delta_kicks}, with an arbitrary degree of disorder. However, in many cases, and for a sufficiently long chain of kicks, the cumulative phase does get randomized and its distribution becomes close to uniform (this can be checked numerically, as discussed in the last section). From Eq.~(\ref{eq:recurrence1}) we can write
\begin{equation}
\ln U_N = \sum_{j=1}^N X_j \, ,
\label{eq:sum_UN}
\end{equation}
and consider that the terms $X_j = \ln(1+2\beta_j) + \ln \left (1 + \gamma_j \,  \cos \Theta_j \right )$ are independent  and identically distributed random variables~\cite{note_sum}. In the large $N$ limit, according to the central-limit theorem, this implies that that $\ln U_N$ has a Gaussian distribution. It only remains to calculate the average and the variance of that distribution. Averaging $X_j$ over the random phase $\Theta_j$~{\cite{note_average}} followed by averaging over $\beta_j$ yields
\begin{equation}
\langle \ln U_N \rangle = N \langle \ln(1+\beta) \rangle \, .
\label{eq:energy_average1}
\end{equation}
Here $\langle ... \rangle$ denotes the full statistical average, over the statistical distributions of $\Theta_j$ and $\beta$ (we have dropped the subscript $j$ since the statistical distribution of $\beta_j$ are taken to be independent of $j$). We have found that in the large $N$ limit $\langle \ln U_N \rangle$ grows linearly with $N$, with a slope $\langle \ln (1+\beta) \rangle$. 
{The linear increase of $\langle \ln U_N \rangle$ for large $N$, or equivalently of $\langle \ln U(t) \rangle$ at long times, is reminiscent of known features of Anderson localization in a random spatially-modulated medium. An analogy can be drawn from two standpoints. The Lyapunov exponent defined in Eq.~(\ref{eq:Lyapunov}) also appears in the one-dimensional Anderson localization problem~\cite{Comtet2013}. In that problem one is interested in the solution to the stationary Schr\"odinger equation, which is different from the (time) Cauchy problem stated in Eq.~(\ref{eq:Helmholtz_k_space}). It can be useful, however, to consider a Cauchy problem for Anderson localization, by fixing the wave function and its spatial derivative at some point in space and then calculating the Lyapunov exponent, that turns out to equal the inverse localization length. We stress that in localization theory, this approach is used as a trick to calculate the localization length, while in the present work the Cauchy problem appears naturally due to physical initial conditions. 
Another analogy, both physical and mathematical, exists between the resistance of a one-dimensional spatially-disordered conductor, and the fraction of backscattered energy $z_N=|D_N^-|^2$ in our problem. More precisely, the resistance $\rho(L)$ of chain with length $L$ is known to grow exponentially with $L$, which is another manifestation of Anderson localization~\cite{Anderson1980}. The backscattered wave energy $z_N$ grows with $N$ in a similar way.}

Calculating the variance of $\ln U_N$ is also possible but one ends up with an integral that cannot be evaluated analytically, unless the disorder is weak. The case of weak disorder is of special interest because it is relevant to experiments (indeed, the modulation of $\varepsilon$ is expected to be very small), and it is amenable to complete analytical treatment.  

{\it Weak disorder.}
We define the weak-disorder regime by the condition $\beta_j \ll 1$. In this case $\langle \ln U_N \rangle = N \langle \beta \rangle$, which follows from (\ref{eq:energy_average1}). This relation shows that $\langle \beta \rangle/2$ is the Lyapunov exponent for weak disorder. To determine the variance, we note that 
$X_j \simeq 2 \beta_j + 2 \sqrt{\beta_j} \cos \Theta_j - 2 \beta_j \cos^2\Theta_j$.
To leading order in $\beta_j$, averaging over the random phase leads to
$\langle X_j^2 \rangle_\Theta \simeq 4 \beta_j \langle \cos^2 \Theta_j \rangle_\Theta = 2 \beta_j$
from which we deduce
\begin{equation}
\mathrm{Var}(\ln U_N) = 2 N \langle \beta \rangle = 2 \langle \ln U_N \rangle \, .
\label{eq:mean_variance}
\end{equation}
We conclude that for large $N$ the wave energy $U_N$ has a log-normal distribution, with mean value and variance satisfying (\ref{eq:mean_variance}). An identical result is known in Anderson localization along weakly disordered chains~\cite{Shapiro1988}, with the resistance being the analog of $U_N$. Our result implies a high degree of universality in one-dimensional wave transport. Not only a universal log-normal distribution is approached for large $N$, but the variance and mean are related by a factor of two, which is a signature of  single parameter scaling.  

The meaning of the large $N$ limit can be clarified. The above treatment is based on Eq.~(\ref{eq:recurrence1}) which, in turn, is based on the assumption $z_j \gg 1$. Initially $z_j$ is very small and gradually grows to reach a value $z_j \sim 1$ after a large number of kicks on the order of $N_c = 1/\langle \beta \rangle$. Thus, the condition for the log-normal distribution and relation (\ref{eq:mean_variance}) is $N \gg N_c$. 

It is also interesting to characterize the intermediate regime $1 \ll N \ll N_c$, in which we can also expect some universal - albeit different - statistical distribution for the wave energy $U_N$. Going back to the general Eq.(\ref{eq:recurrence_general}), which has no restriction on the value of $z_j$, we set there $z_j, z_{j-1} \ll 1$ as well as $\beta_j \ll 1$. This leads to the recursion relation
\begin{equation}
z_j = z_{j-1} + \beta_j + 2 \sqrt{\beta_j z_{j-1}} \cos \Theta_j \ \ (j \ll N_c) \, .
\label{eq:recurrence_small_z}
\end{equation}
As before, we assume that the phase $\Theta_j$ is completely random or, at least, that it gets randomized after some number of kicks $j_0 \ll N_c$. Next, we raise Eq.~(\ref{eq:recurrence_small_z}) to power $n$ and average first over $\Theta_j$, and then over some arbitrary distribution of $\beta_j$, keeping only the leading (linear) terms in $\beta_j$. {This enables us to express the $n^{th}$ moment of $z_j$ in the form (see Supplemental Material~\cite{SM})}
$\langle z_j^n \rangle = (n!) \langle z_j \rangle^n$ for $j_0 \ll j \ll N_c$,
with $\langle z_j \rangle = j \langle \beta \rangle$. This implies that after a sequence of $N$ kicks, $z_N$ follows a negative exponential (or Rayleigh) distribution:
\begin{equation}
P(z_N) = (N \langle \beta \rangle)^{-1} \exp[-z_N/(N \langle \beta \rangle)] \ \ (1 \ll N \ll N_c) \, .
\label{eq:negative_exp}
\end{equation}
We conclude that the energy $U_N=1+2z_N$ in this regime has negative exponential distribution, and that the average energy $\langle U_N \rangle = 1+2 N \langle \beta \rangle$ grows linearly with the number of kicks.

It is possible to treat both regimes of short and long chains, or equivalently short and long times, using a more formal approach based on a variant of a Fokker-Planck equation, sometimes referred to as Melnikov's equation~\cite{Melnikov_etal}. To proceed, we start with the basic recursion relation (\ref{eq:recurrence_general}) for the variable $z$ and transform it into a recursion relation for the distribution $P_j(z)$ for that variable after $j$ kicks {(the derivation, given in Ref.~\cite{Melnikov_etal}, is summarized in the Supplemental Materiel~\cite{SM})}. In the weak-disorder regime $\langle \beta \rangle \ll 1$, this equation is
\begin{equation}
P_j(z) = P_{j-1}(z) + \langle \beta \rangle \, \frac{\partial}{\partial z} \left [(z+z^2)  \frac{\partial P_{j-1}}{\partial z}(z) \right ] \, .
\end{equation}
Next, we transform the discrete time steps $t_j$ into the continuous time $t$, by using the average time interval $\Delta t = \langle t_j - t_{j-1} \rangle$. This leads to
\begin{equation}
\frac{\partial P_t}{\partial t}(z) = \alpha \, \frac{\partial}{\partial z} \left [ (z+z^2)  \frac{\partial P_t}{\partial z}(z) \right ] \, ,
\label{eq:Pj_continuous}
\end{equation}
with $\alpha = \langle \beta \rangle/\Delta t$. 
In principle, Eq.~(\ref{eq:Pj_continuous}) should be solved with an initial condition $P_{t=0}(z)$. Actually, the precise shape of the initial distribution is rapidly forgotten and a universal function of $z$ (with $\alpha$ as a single parameter) is approached as time elapses. There are two distinct regimes, which can be separated using the critical time $\tau_c = 1/\alpha$, which is the counterpart of $N_c$ in the continuous time picture. At short times $t \ll \tau_c$, $z$ remains small and we can neglect the $z^2$ term in (\ref{eq:Pj_continuous}) to obtain
\begin{equation}
P_t(z) = (\alpha t)^{-1} \exp[-z/(\alpha t)] \ \ (t \ll \tau_c) \, .
\end{equation}
We find that $z$ follows a negative exponential distribution, identical to Eq.~(\ref{eq:negative_exp}), but for continuous time. In the opposite limit $t \gg \tau_c$, the $z$ term in (\ref{eq:Pj_continuous}) can be neglected, and a log-normal distribution for $z$ is obtained:
\begin{equation}
P_t(z) = (z \sqrt{4\pi\alpha t})^{-1} \exp[-(\ln z - \alpha t)^2/(4 \alpha t)] \ \ (t \gg \tau_c) \, .
\end{equation}
This long-time statistics is in agreement with that obtained previously for a discrete chain of kicks in the limit $N \gg N_c$. 

{\it Numerical results.}
In order to support and illustrate the theoretical analyses, we have carried out numerical simulations, using the $\delta$-kicks model defined in Fig.~\ref{fig:delta_kicks}. The transfer matrix in this case takes the form (\ref{eq:Mj_general}), with coefficients given by Eq.~(\ref{eq:ajbj_kicks}). In the simulations $u_j$ and $\theta_j$ are taken to be uniformly distributed random variables, with $u_j\in[0,0.05]$ (corresponding to weak disorder) and $\theta_j\in [0,2\pi]$. By performing products of transfer matrices, we simulate a random chain of $N$ kicks, and calculate numerically the energy $U_N$. Doing this for many realizations of disorder ({\it i.e.} of $u_j$ and $\theta_j$), we can compute the statistical distributions of $U_N$ or $\ln U_N$, and compare the numerical results with the theoretical predictions. 

\begin{figure}[h]
     \begin{center}
     \includegraphics[width=8cm]{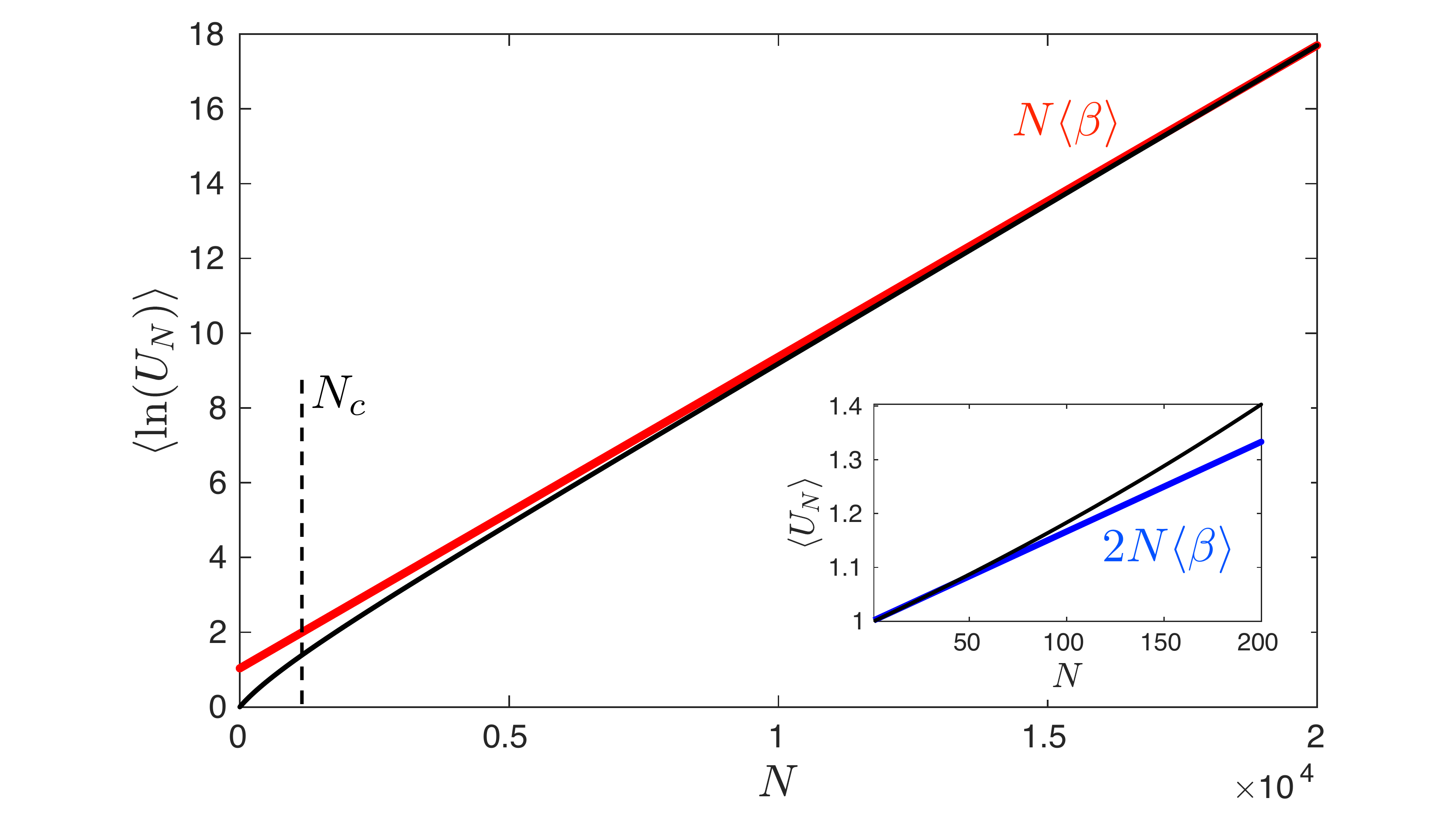}
     \end{center}    
     \caption{\label{fig:lnU} Numerical simuation of $\langle \ln U_N \rangle$ versus the number of kicks $N$ (black line). For $N \gg N_c$ ($N_c \simeq 1200$ here) we observe a linear growth with a slope $\langle \beta \rangle$ (indicated by the red line). For $1 \ll N \ll N_c$, we observe a regime where $\langle U_N \rangle \sim 1+2N \langle \beta \rangle$, as shown in the inset. The numerical simulations confirm the theoretical predictions. A $\delta$-kicks model as in Fig.~\ref{fig:delta_kicks} is used in the simulations, with $u_j$ and $\theta_j$ uniformly distributed in $[0,0.05]$ and $[0,2\pi]$.
            }
\end{figure}

Focusing first on the average energy, we show in Fig.~\ref{fig:lnU} a plot of $\langle \ln U_N \rangle$ versus the number of kicks $N$. For $N \gg N_c$, with $N_c=1/\langle \beta \rangle \simeq 1200$ here, we find that $\langle \ln U_N \rangle$ grows linearly with $N$, with a slope $\langle \beta \rangle \simeq 8.3 \times 10^{-4}$ coinciding with that predicted theoretically, as indicated by the straight line. In the region $1 \ll N \ll N_c$, we observe a regime in which $\langle U_N \rangle \sim 1+ 2 N \langle \beta \rangle$, also predicted theoretically. An enlargement of this intermediate regime is shown in the inset. Although not shown for brevity, we have checked numerically that the condition of the random phase model is satisfied as soon as $N \gg 1$, and that the condition (\ref{eq:condition2}) is satisfied for $N \gg N_c$. 

\begin{figure}[h]
     \begin{center}
     \includegraphics[width=8cm]{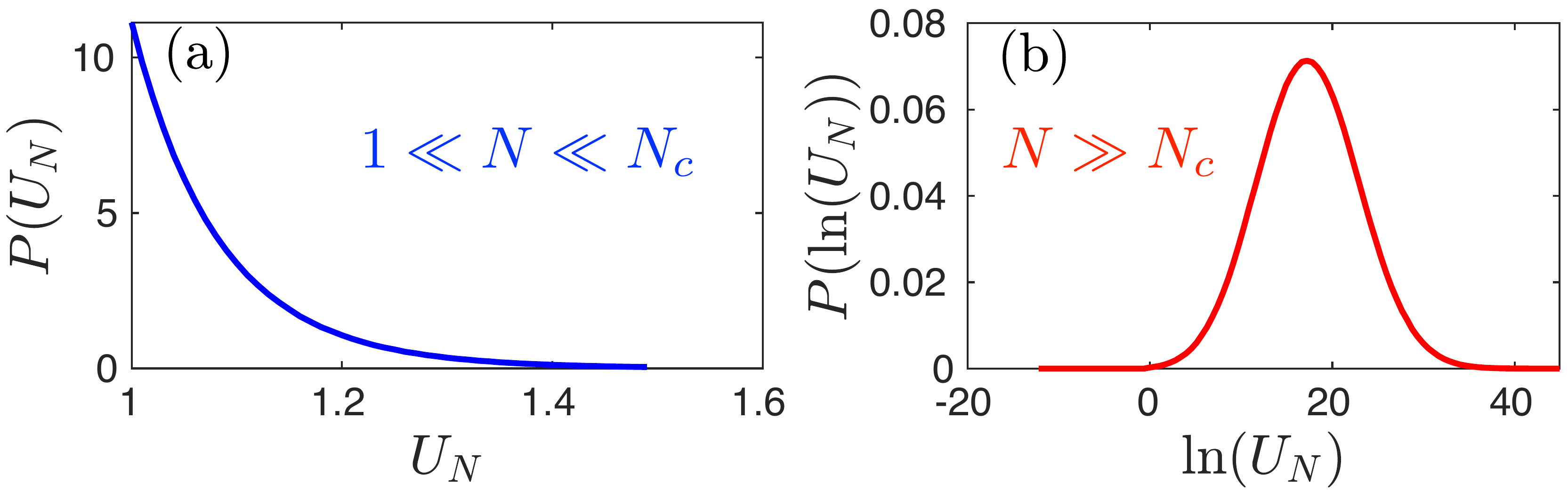}
     \end{center}    
     \caption{\label{fig:stat_U} (a): Statistical distribution of $U_N$ in the intermediate regime $1 \ll N \ll N_C$ ($N=50$). A negative exponential distribution is observed, in full agreement with theory. 
          (b): Statistical distribution of $\ln U_N$ in the asymptotic regime $N \gg N_c$ ($N=2 \times 10^4$). A Gaussian distribution is observed, also in agreement with theory. Same model as in Fig.~\ref{fig:lnU}.}
\end{figure}

The numerical simulation also permits a computation of the full statistical distribution of $U_N$. The distributions in the intermediate regime $1 \ll N \ll N_c$ and in the large $N$ limit $N \gg N_c$ are shown in Figs.~\ref{fig:stat_U} (a) and (b). In the intermediate regime, we find that the energy $U_N$ follows a negative exponential law, with average value $\langle U_N \rangle = 1+2N \langle \beta \rangle$. For $N \gg N_c$, we find that the distribution of $U_N$ is log-normal ($\ln U_N$ is Gaussian), with $\langle \ln U_N \rangle = N \langle \beta \rangle$ and $\mathrm{Var}(\ln U_N) = 2 \langle \ln U_N \rangle$. The calculated statistical distributions perfectly match the theoretical predictions, and confirm the universal character and single-parameter scaling of wave transport in randomly time-varying homogeneous and isotropic media, in the regime of weak disorder.

{\it Conclusion.}
In summary, we have presented a general model for wave propagation in a random time-dependent medium, and demonstrated the existence of universal statistical distributions of the wave energy $U$. We proved that, after a sufficiently long time, $U$ approaches a log-normal distribution with $\langle \ln U \rangle \sim t$, {in agreement with well-established results in one-dimensional transport.} In the weak-modulation regime, a full analytic theory was developed, which reveals two distinct regimes: For time smaller than some crossover time $\tau_c$, the energy distribution follows a negative exponential (Rayleigh) distribution, while for times $t \gg \tau_c$ the distribution crosses over to a log-normal law. {The intermediate regime for $t \ll \tau_c$ might be relevant to experiments in which the long-time regime could be difficult to reach.} The theory, in perfect agreement with numerical simulations, lays some foundation in the emerging topic of waves in disordered temporal media, with expected outcomes in the control of various kinds of waves.

\begin{acknowledgments}
We thank A. Piette for his help in the early stage of the project, and E.~Fort and M. Fink for stimulating discussions. Work supported by LABEX WIFI (Laboratory of Excellence within the French Program Investments for the Future) under references ANR-10- LABX-24 and ANR-10-IDEX-0001-02 PSL*.
\end{acknowledgments}


\newpage

\section*{Supplemental Material}

\section{Random barriers model}

\begin{figure}[h]
     \begin{center}
     \includegraphics[width=7.5cm]{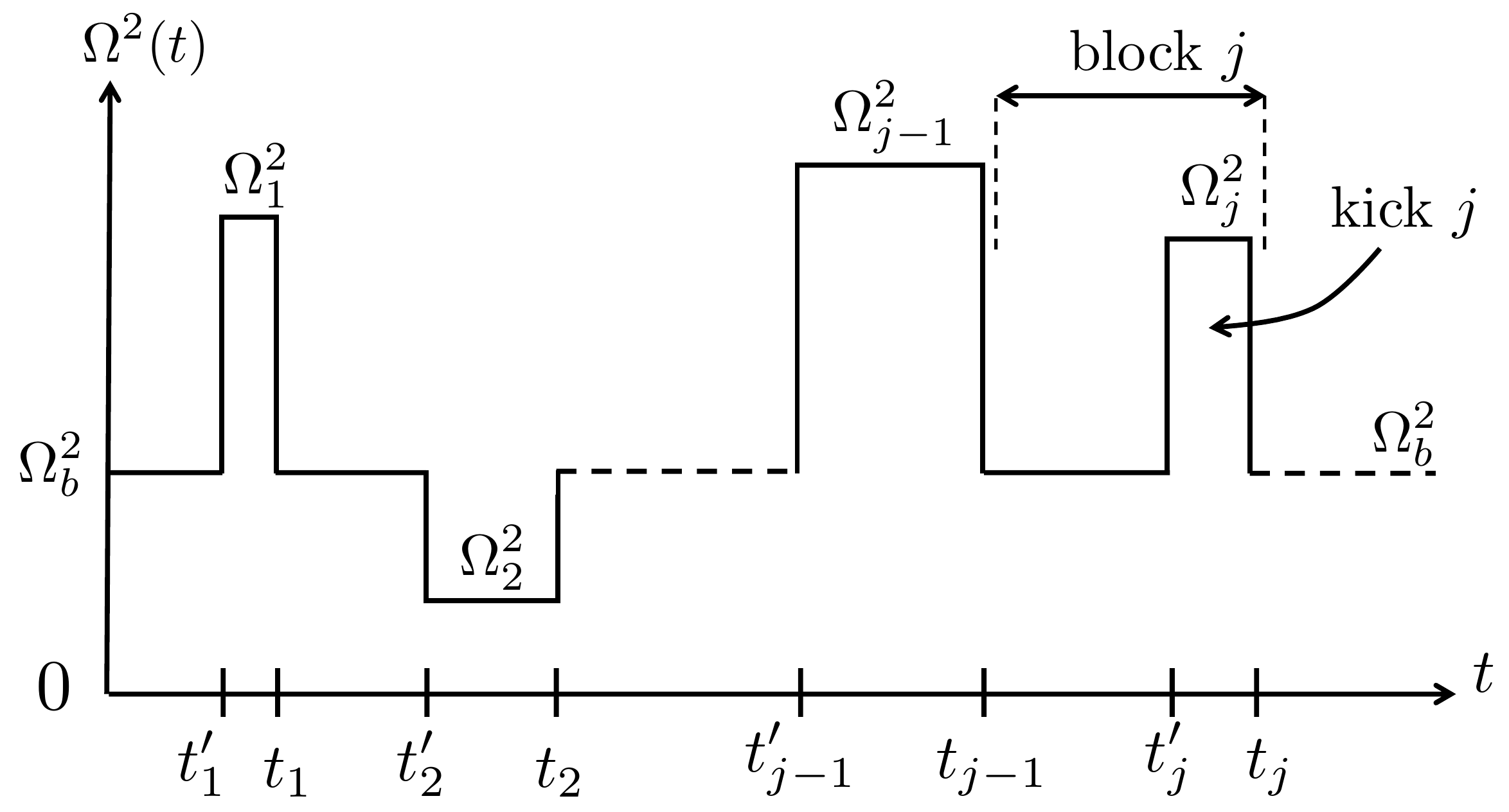}
     \end{center}    
     \caption{\label{fig:random_barriers} Random chain of rectangular barriers, each of them referred to as a ``kick''. The barrier heights $\Omega^2_j$ and the times $t'_j$ and $t_j$ are random variables. After each kick the medium recovers the background value $\Omega^2_b$.}
\end{figure}

In this section we describe a model for a time-disordered homogeneous medium that could be used as an alternative to the $\delta$-kicks model presented in the main text. $\Omega^2(t)$ is considered to be a chain of rectangular barriers, as represented in Fig.~\ref{fig:random_barriers}. Each barrier is denoted as a kick, with the times $t'_j$  and $t_j$ defining the onset and end of kick number $j$. The times $t'_j$  and $t_j$, as well as the kick strengths (barrier heights) $\Omega^2_j$ are random variables.

The transfer matrix $\Mg_j$ of a single block $j$ takes the form
\begin{equation}
\Mg_j
=
\begin{bmatrix}
a_j & b_j \\
b_j^* & a_j^*
\end{bmatrix}
\label{eq:Mj_general2}
\end{equation}
with
\begin{eqnarray}\label{a_coeff}
   a_j &=& T_j T'_j  \exp[-i(\theta_j+\phi_j)] + R_j R'_j \exp[-i(\theta_j-\phi_j)]
\\\label{b_coeff}
   b_j &=& T_j R'_j  \exp[i(\theta_j-\phi_j)] + R_j T'_j \exp[i(\theta_j+\phi_j)] \, .
\end{eqnarray}
Here, $T_j = (\Omega_b+\Omega_j)/(2\Omega_b)$, $T'_j = (\Omega_b+\Omega_j)/(2\Omega_j)$, $R_j = (\Omega_b-\Omega_j)/(2\Omega_b)$ and $R'_j = (\Omega_j-\Omega_b)/(2\Omega_j)$ are time-domain transmission and reflection coefficients, that can be deduced from the continuity of the field and its time derivative at the times $t=t^\prime_j$ and $t=t_j$. The phases $\phi_j=\Omega_j(t_j-t'_j)$ and $\theta_j=\Omega_b(t'_j-t_{j-1})$ correspond to free propagation during the kick, and free propagation between two succesive kicks, respectively. 

We note that the transfer matrix takes the same form as Eq.~(6) in the main text, but with different matrix elements. It can be verified that the matrix elements also satisfy Eqs.~(8) and (9) of the main text, which are very general and independent of the model of disorder (see~Ref.~\cite{MelloBook} for a general derivation of these properties).

\section{Transfer matrix for $\delta$-kicks}

In this section we deduce the expression of the transfer matrix for the $\delta$-kicks model described in Fig.~1 of the
main text, starting from the random barriers model decribed in the previous section. For kick number $j$ we take a barrier with duration $\delta t$ and amplitude such that
\begin{equation}
   t_j-t'_j=\delta t,
   \quad
   \Omega_j^2-\Omega_b^2=v_j/\delta t.
\end{equation}
Taking the limit $\delta t\to 0$ leads to $\Omega^2(t)=\Omega_b^2+\sum_j v_j\delta(t-t_j)$ which corresponds to the
$\delta$-kicks model with $v_j$ the amplitude of kick number $j$. Taking the same limit for the coefficients of the transfer matrix $\Mg_j$ in
Eqs.~(\ref{a_coeff}) and (\ref{b_coeff}) leads to
\begin{eqnarray}
   a_j &=& [1-iv_j/(2\Omega_b)]\exp(-i\theta_j),
\\
   b_j &=& -iv_j/(2\Omega_b)\exp(i\theta_j).
\end{eqnarray}
These expressions correspond to Eq.~(7) of the main text.

\section{Derivation of the moment relation leading to Eq.~(20)}

In this section we derive the moment relation $\langle z_j^n \rangle = (n!) \langle z_j \rangle^n$ that is used as a step
in the derivation of Eq.~(20) in the main text. We start by raising Eq.~(19) of the main text to the power $n$. Keeping terms of order $\sqrt{\beta_j}$ and $\beta_j$, we obtain
\begin{multline}
   z_j^n=z_{j-1}^n+n\beta_jz_{j-1}^{n-1}+2nz_{j-1}^{n-1}\sqrt{\beta_jz_{j-1}}\cos(\Theta_j)
\\
      +2n(n-1)z_{j-1}^{n-1}\beta_j\cos(\Theta_j)^2+\operatorname{O}(\beta_j^{3/2}).
\end{multline}
We perform an average over $\Theta$, assuming that it is fully randomized after a sufficiently large number of kicks (this hypothesis is discussed in the main text) and independent of $\beta_j$, and a subsequent average over an arbitrary distribution for $\beta_j$. This leads to the following recursion relation for the moments of $z$
\begin{equation}\label{recurrence}
   \langle z_j^n\rangle = \langle z_{j-1}^n\rangle
      +n^2\langle\beta\rangle\langle z_{j-1}^{n-1}\rangle  \, ,
\end{equation}
with the first moment given by $\langle z_j\rangle=j\langle\beta\rangle$. We now assume that the moment of order $n$ is
\begin{equation}
   \langle z_j^n\rangle = n! \langle z_j\rangle^n = n! j^n\langle\beta\rangle^n.
\end{equation}
Inserting the above relation in Eq.~(\ref{recurrence}), it is easy to see that it is satisfied up to terms of order $1/j^2$. This concludes the derivation of the relation $\langle z_j^n\rangle = n! \langle z_j\rangle^n$, valid for large enough $j$.

\section{Derivation of Melnikov's equation}

In this section we derive Eq.~(21) of the main text. The derivation can be found in Ref.~[S1], and we summarize the main steps here. We starts with the basic recursion relation for the variable $z$ [Eq.~(12) in the main text], that can be rewritten as
\begin{multline}
   z_j=z_{j-1}+\beta_j(1+2z_{j-1})
\\
      +2\sqrt{\beta_j(1+\beta_j)z_{j-1}(1+z_{j-1})}\cos\Theta_j.
\end{multline}
Considering as above that $\Theta_j$ is fully randomized, the above relation can be transformed into a recursion relation for
the probability density $P_j(z)$, which takes the form
\begin{multline}\label{distribution}
   P_j(z) = \int_0^{\infty}f(\beta)d\beta\int_0^{2\pi}\frac{d\Theta}{2\pi}
\\\times
      P_{j-1}\left(z+\beta(1+2z)-2\sqrt{\beta(1+\beta)z(1+z)}\cos\Theta\right) \, ,
\end{multline}
where $f(\beta)$ is the probability density asociated to the random variable $\beta$. For weak disorder such that
$\beta\ll 1$, we can perform a second order Taylor expansion for $P_{j-1}$, which leads to
\begin{multline}
   P_{j-1}(z+\epsilon)= P_{j-1}(z)
      +\epsilon\frac{\partial P_{j-1}(z)}{\partial z}
      +\frac{\epsilon^2}{2}\frac{\partial^2 P_{j-1}(z)}{\partial z^2}
\\
      +\operatorname{O}(\epsilon^3) \, ,
\end{multline}
where $\epsilon=\beta(1+2z)-2\sqrt{\beta(1+\beta)z(1+z)}\cos\Theta$.
Substituting in Eq.~(\ref{distribution}), we obtain
\begin{multline}
   P_j(z) = P_{j-1}(z)+\langle\beta\rangle(1+2z)\frac{\partial P_{j-1}(z)}{\partial z}
\\
      +\langle\beta\rangle z(1+z)\frac{\partial^2 P_{j-1}(z)}{\partial z^2}
      +\operatorname{O}(\beta^2) \, .
\end{multline}
This can be factorized in the form
\begin{equation}
   P_j(z) = P_{j-1}(z) + \langle \beta \rangle \frac{\partial}{\partial z}
      \left [(z+z^2)  \frac{\partial P_{j-1}}{\partial z}(z) \right ] \, ,
\end{equation}
where terms beyond first order have been neglected. This is Eq.~(21) in the main text, which in the continuous
limit leads to Melnikov's equation~(22). 

\subsection*{References}
[S1] P.A. Mello and N. Kumar, {\it Quantum Transport in Mesoscopic Systems: Complexity and Statistical Fluctuations. A Maximum Entropy Viewpoint} (Oxford University Press, 2004).


\begin{thebibliography}{40}
\bibitem{note_sum}
Since phase randomization is established after a certain number of kicks, the sum in Eq.~(\ref{eq:sum_UN}) should start from some model-dependent value $j_0 > 1$. This, however, makes no difference in the large $N$ limit.
\end{thebibliography}
\end{document}